\documentstyle[12pt]{article}
\date{}
\def\mc{\multicolumn}
\def\be{\begin{equation}}
\def\ee{\end{equation}}
\def\ben{\begin{eqnarray}}
\def\een{\end{eqnarray}}
\def\bena{\begin{eqnarray*}}
\def\eena{\end{eqnarray*}}
\def\bdes{\begin{description}}
\def\edes{\end{description}}
\def\benum{\begin{enumerate}}
\def\eenum{\end{enumerate}}

\def\ra{\rightarrow}

        \def\g{\gamma}  
    \def\t{\tau}    
\def\p{\psi}               
           \def\l{\lambda}
\def\L{\Lambda}  \def\pa{\partial}
\title{%
\vskip-7em\hfill {\small \begin{tabular}[t]{l}
                  \rule{0ex}{1ex} CERN-TH.7028/93 \\[.0ex]
                  \rule{0ex}{1ex} MS-TPI-93-07    \\[.0ex]
                  \rule{0ex}{1ex} October 1993
\end{tabular} \break\vskip3em  }
\Large\bf
Monte Carlo Simulation of 2-D Quantum Gravity
as Open Dynamically Triangulate Random Surfaces}
\author{\vbox{\vspace{0mm}}\\[-4mm]%
E. Adi$^1$, M. Hasenbusch$^2$, M. Marcu$^3$,  \\
E. Pazy$^1$, K. Pinn$^4$, and S. Solomon$^1$ \\[6mm]
\small $^1\,$ Racah Institute of Physics           \\[-2mm]
\small Hebrew University, 91904 Jerusalem, Israel  \\[-2mm]
\small eti@vms.huji.ac.il                          \\[-2mm]
\small pazy@vms.huji.ac.il                         \\[-2mm]
\small sorin@vms.huji.ac.il                        \\[1mm]
\small $^2\,$CERN Theory Division                  \\[-2mm]
\small CH-1211 Gen\`eve 23, Switzerland            \\[-2mm]
\small hasenbus@surya11.cern.ch                    \\[1mm]
\small $^3\,$School of Physics and Astronomy                 \\[-2mm]
\small Raymond and Beverly Sackler Faculty of Exact Sciences \\[-2mm]
\small Tel Aviv University, 69978 Tel Aviv, Israel           \\[-2mm]
\small marcu@vm.tau.ac.il                                    \\[1mm]
\small $^4\,$Institut f\"ur Theoretische Physik I,
       Universit\"at M\"unster                               \\[-2mm]
\small Wilhelm-Klemm-Str.\ 9, D-48149 M\"unster, Germany      \\[-2mm]
\small pinn@yukawa.uni-muenster.de                           \\[5mm]
}
\begin{document}
\maketitle %%\vfill
\thispagestyle{empty}
\begin{abstract}\noindent \normalsize \vbox{\vspace{0mm}}\\[-4mm]
We describe a Monte Carlo procedure for the simulation
of dynamically triangulate random surfaces with a boundary
(topology of a disk). The algorithm
keeps the total number of triangles fixed, while the length
of the boundary is allowed to fluctuate.
The algorithm works in the presence of matter fields.

We here present results for the pure gravity case.
The algorithm reproduces the theoretical expectations.
\end{abstract}
\setcounter{page}{0}
\newpage

\section{Introduction}
\label{SECintro}
The quantization of the 2-dimensional Einstein-Hilbert
theory of gravity presents
conceptual and technical difficulties \cite{problems}.
In the Euclidean path integral formulation, one has
to integrate over all metrics (modulo diffeomorphisms)
of a Riemannian manifold with fixed topology.
One way of making sense out of this ill defined integral
is to approximate
the manifolds by triangulations. The integration over all
metrics is then replaced by a summation over all
triangulations.

The approach to quantum gravity via Dynamically Triangulate
Random Surfaces (DTRS) opens the way for the use of
Monte Carlo methods which were proven to be very
useful, especially if it comes to the inclusion of
matter fields \cite{success}.

In this letter, we present a Monte Carlo procedure for the
simulation of DTRS that have a boundary.

The importance of the boundary comes from the fact that
the estimation of the probability distribution of its length
corresponds to a measurement of the wave function of the 2D universe.
Its theoretical treatment turned out to be problematic.
With only the Hilbert-Einstein term in the action,
2-D quantum gravity is non-renormalizable by
naive power counting and is known to suffer from
ultraviolet divergences.
Furthermore, when conformal field theory
is coupled to 2-D gravity, a naive counting
yields ``$-1$'' degrees of freedom \cite{elitzur}.
This is believed to reflect a
non-normalizable
universe wave function solution of the Wheeler-de Witt equations.
Matrix model results indicate too that
the wave function might be
non-normalizable at small areas.
These considerations led us
to a numerical study of open random surfaces (surfaces with a boundary).
The hope is that adding extra boundary terms
(or additional matter fields) to the action
might lead to a normalizable wave function.

It is therefore useful to have a Monte Carlo algorithm for the
simulation of open random surfaces, not only to understand
the nature of a possible normalizable wave function but also
as a tool to study the intrinsic geometry of open surfaces.

A Monte Carlo procedure for the simulation of closed surfaces
with a fixed number of triangles,
using the (2,2) Alexander move (to be called ``bulk flip'' later)
was studied by Kazakov et al.\ \cite{kazakov}.
A generalization of this type of algorithm to the case of
a surface with a boundary with fluctuating length turns out to
be nontrivial. The main problem is to ensure ergodicity and
detailed balance (i.e.\ that each triangulation contributes
with equal probability to the partition function).

This letter is organized as follows.
The model is introduced in section~\ref{SECmodel}.
Section~\ref{SECalgorithm} is devoted to the description
of the algorithm. In section~\ref{SECresults} we
present our results and compare them with the theoretical
values.

In the present paper we limit ourselves
to the pure gravity case where
most of our results can be obtained also analytically.
The inclusion of the matter fields (possibly
with central charge $c > 1$) where no analytical results
are available is straightforward algorithmically, but its
theoretical diagnostics and interpretation is beyond the
framework of the present letter.
\section {The model}
\label{SECmodel}
The Euclidean Einstein-Hilbert action is
%2
\be\label{action}
S_{E}[g]=\frac{1}{16 \pi G_{N}} \int_{M} d^{2}x
\, \sqrt{|g|} \, (-R + 2 \L) \, .
\ee
Here $G_{N}$ is Newton's constant, $R$ is the intrinsic curvature, and $\L$ is
the cosmological constant, in units of $\hbar=c=1$. $M$ is a
Riemannian space time manifold.

In the continuum quantized version (for the sake of completeness
we include matter fields $X$ here) one considers the partition
function
%3
\be\label{continuum}
Z=\int \frac{DgDX}{{\rm vol(Diff)}} \,
\exp\left( -S_{E}[g] - S_{m}[X,g] \right) \, .
\ee
The discrete analog of eq.~(\ref{continuum}) without matter fields is
%4
\be\label{discrete}
Z=\sum_{\{\tau\}} \frac {1}{s(\tau)} \exp\left(a_{0}N_{0} +
a_{1}N_{1} + a_{2}N_{2}\right) \, .
\ee
The sum is over all {\em oriented} triangulations $\tau$ of the
surface. $N_{0},N_{1},N_{2}$ are the number of vertices, links
and triangles of $\tau$, respectively,
and $s(\tau)$ is the order of the symmetry group
of the triangulation $\tau$.
$a_{0},a_{1},a_{2}$ are free parameters.
The factor $\frac{1}{s(\tau)}$ can be viewed as what remains from the
$\frac{1}{{\rm vol(Diff)}}$ factor in eq.~(\ref{discrete}).

For open surfaces the action in
eq.~(\ref{continuum}) picks up two additional terms
(as in the Liouville theory),
%5
\be
S \ra S + \frac{1}{16 \pi G_{N}} \oint_{\pa M} d\hat{s} \,
\sqrt{\tilde{h}} \, (-k + 4\Xi) \, ,
\ee
where $\tilde{h}$ is the induced boundary metric,
$k$ is the extrinsic curvature on the boundary,
and $\Xi$ is the boundary cosmological constant.

The Einstein-Hilbert action is a topological invariant, since the
2-D oriented manifolds obey the Gauss-Bonnet theorem
(with the correction arising from the boundary term):
%6
\be
\frac{1}{4\pi}( \int_{M} d^{2}x \, \sqrt{|g|} \, R
+ \oint_{\pa M} \, \sqrt{\tilde{h}} \, d\hat{s} \, k) = \chi \, .
\ee
Here $\chi=2 - 2h - b$ is the Euler characteristic.
$h$ is the number of handles,
and $b$ is the number of boundaries. Nevertheless some non-trivial quantum
theory exists, mainly due to conformal anomalies \cite{polyakov}.

{}From the topological relation $\chi = N_{0} - N_{1} + N_{2}$,
and the relation
for a connected triangular lattice $2 N_{1} = 3 N_{2} + L$,
where $L$ is the boundary length,
the discrete action in eq.~(\ref{discrete})
can be written as
%7
\be
a_{0}N_{0} +a_{1}N_{1} + a_{2}N_{2}  =  -\l N_{2} - \xi L + \g \chi \, ,
\ee
with
\begin{eqnarray}
\l  & = & -(\frac {a_{0}}{2} + \frac{3a_{1}}{2} + a_{2}) \nonumber \\
\xi & = & -\frac{a_{0}+a_{1}}{2}  \nonumber   \\
\g  & = & a_{0} \, .
\end{eqnarray}
The action depends on three independent parameters: $\l$ and $\xi$,
that we identify with the cosmological constants $\L$ and $\Xi$, and $\g$,
that we identify with $1/G_{N}$. Actually as we fix $\chi$, since we
identify $\g$ with a constant term,
we expect the partition function to look like
%8
\be
Z=\sum_{\{\tau\}} \frac {1}{s(\tau)} \,  e^{-\l N} \,
e^{-\xi L} \, = \sum_{N,L} Z(L,N) \, e^{-\l N} e^{-\xi L} \, .
\ee
Here and in the following we have identified $N \equiv N_{2}$.

In our simulations to be specified below one estimates $Z(N,L)$
 which is the combinatorial
weight of a surface with $N$ triangles and $L$ external edges.

We consider open connected triangulate
surfaces without 1- and 2-loops.\footnote{
Given a triangulate surface $S$, a $p$-loop is a set of $p$ distinct edges
in $S$ which form a loop on $S$}
In the language of the dual
lattice, where a triangle becomes a point, an edge a segment and
a vertex a polygon, every surface $S$ is in
one-to-one correspondence with a
connected planar diagram of the $\phi^{3}$ theory without tadpoles and
self-energy.

One of our aims is to find the wave function $\p(L)$,
which is the probability amplitude to have a boundary length $L$,
for a {\em fixed} number of triangles $N$.
Our Monte Carlo procedure samples the configuration space with no
a priori weight (each configuration occurs with the same
probability). We can therefore determine the
probability that a configuration has boundary length $L$
by just making a simple histogram, cf.\ section~\ref{SECresults}.

As a by-product of the simulation one gets knowledge about the
intrinsic geometry of the surface.
Using the simulation results, we are able to
picture and characterize the typical surfaces.
\section{Algorithm}
\label{SECalgorithm}

We consider a connected,
open, triangulate lattice, with the topology of a disk.
By definition, all edges have equal length.
We have to distinguish three types of triangles:
\begin{itemize}
  \item {\em type-0} triangles have {\em none}
   of their three edges on the boundary of the surface (bulk triangles),
  \item {\em type-1} triangles share exactly {\em one}
   edge with the boundary,
  \item {\em type-2} triangles have {\em two} edges
   in common with the boundary.
\end{itemize}
Our Monte Carlo algorithm generates a start configuration
that consists of $N$ triangles and that has a certain
boundary length $L_0$. Then a number of sweeps is done.
Each sweep consists of a sequence of elementary flips
done on the configuration:

\begin{itemize}
\item {\em bulk flips} change the internal
geometry but maintain a fixed boundary length.
\item {\em boundary flips} change the boundary length $L$. There
are two kinds of boundary flips:
  \begin{itemize}
   \item {\em type-1 boundary flips} are performed on
   {\em type-1} triangles and increase the boundary length by 2.
   \item {\em type-2 boundary flips}
   act on {\em type-2} triangles and decrease the boundary length by 2.
\end{itemize}
\end{itemize}

Because of the relation $E=\frac{3N+L}{2}$ for a connected triangular
lattice, where $E$ is the (integer) number of edges in the configuration,
we deduce that for fixed $N$,
the boundary length $L$ may be incremented by even values only.

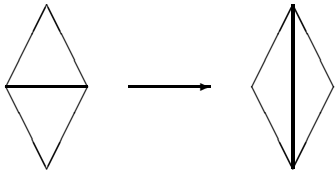
\begin{figure}[h]
\begin{center}
\setlength{\unitlength}{3ex}
\begin{picture}(15,7)(0,0)
 \put(4,3)
 {
  \begin{picture}(8,5)(0,0)
   \put(-1,0){\line( 1, 2){1}}
   \put(-1,0){\line( 1,-2){1}}
   \put( 1,0){\line(-1, 2){1}}
   \put( 1,0){\line(-1,-2){1}}
   \thicklines
   \put(-1,0){\line(1,0){2}}
   \thinlines
   \put( 2,0){\vector(1,0){2}}
  \end{picture}
 }
 \put(10,3)
 {
  \begin{picture}(8,5)(0,0)
   \put(-1,0){\line( 1, 2){1}}
   \put(-1,0){\line( 1,-2){1}}
   \put( 1,0){\line(-1, 2){1}}
   \put( 1,0){\line(-1,-2){1}}
   \thicklines
   \put(0,-2){\line(0,1){4}}
   \thinlines
  \end{picture}
 }
\end{picture}
  \parbox[t]{.85\textwidth}
  {
  \caption[The bulk flip]
  {\label{figBulkflip}
  The {\em bulk flip}}
  }
\end{center}
\end{figure}

\begin{figure}
\begin{center}
\setlength{\unitlength}{3ex}

\begin{picture}(10,6)(0,0)
 \put(-2,3)
 {
  \begin{picture}(8,5)(0,0)
   \put( 0, -2){(a)}
   \put(-2, 2){\line( 1,-1){2}}
   \put(-2, 2){\line( 0,-1){3}}
   \put( 2, 1){\line(-1, 1){1}}
   \thicklines
   \put(-1, 1){\line( 1, 0){3}}
   \put(-1, 1){\line( 2, 1){2}}
   \put( 2, 1){\line(-1, 1){1}}
   \thinlines
   \put( 0, 0){\line( 2, 1){2}}
   \put( 2, 1){\line( 1, 1){1}}
   \put( 2, 1){\line( 0,-1){2}}
   \put( 2,-1){\line( 1, 3){1}}
   \put(-2,-1){\line( 1, 2){1}}
   \multiput(-1,1)(0.2,0.05){20}{\makebox(0,0){.}}
   \multiput( 2,1)(-0.2,0.05){20}{\makebox(0,0){.}}
  \end{picture}
 }
 \put(5,3)
 {
  \begin{picture}(8,5)(0,0)
   \put( 0, -2){(b)}
   \put(-2, 2){\line( 1,-1){2}}
   \put(-2, 2){\line( 0,-1){3}}
   \thicklines
   \put(-1, 1){\line( 1, 0){3}}
   \thinlines
   \put( 0, 0){\line( 2, 1){2}}
   \put( 2, 1){\line( 1, 1){1}}
   \put( 2, 1){\line( 0,-1){2}}
   \put( 2,-1){\line( 1, 3){1}}
   \put(-2,-1){\line( 1, 2){1}}
   \multiput(-1,1)(0.2,0.05){20}{\makebox(0,0){.}}
   \multiput( 2,1)(-0.2,0.05){20}{\makebox(0,0){.}}
  \end{picture}
 }
 \put(12,3)
 {
  \begin{picture}(8,5)(0,0)
   \put( 0, -2){(c)}
   \put(-2, 2){\line( 1,-1){2}}
   \put(-2, 2){\line( 0,-1){3}}
   \thicklines
   \put(-1, 1){\line( 1, 0){3}}
   \thinlines
   \put( 0, 0){\line( 2, 1){2}}
   \put( 2, 1){\line( 1, 1){1}}
   \put( 2, 1){\line( 0,-1){2}}
   \put( 2,-1){\line( 1, 3){1}}
   \put(-2,-1){\line( 1, 2){1}}
   \put( 2, 1){\line(-4, 1){4}}
  \end{picture}
 }
\end{picture}
  \parbox[t]{.85\textwidth}
  {
  \caption[The {\em type-2} boundary flip]
  {\label{figFlipType-2}
  How the {\em type-2 boundary flip} works: (a) Two virtual
  edges are added to the configuration. (b) The {\em type-2}
  triangle is removed. (c) One of the virtual links is removed,
  the other is made real.}
  }
\end{center}
\end{figure}
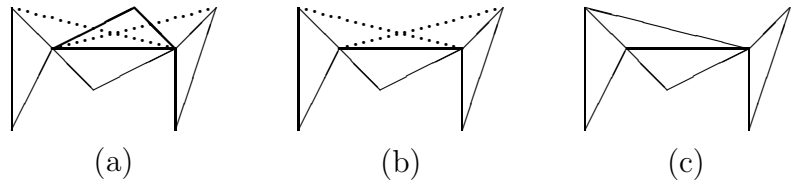

Let us characterize in detail each of the three flip operations:
\begin{itemize}
\item {\em Bulk flips}
(see fig.~\ref{figBulkflip}):
\newline
These are the basic moves also used in the simulation
of closed DTRS \cite{closed-DTRS}, see fig.~\ref{figBulkflip}.

\item {\em Type-1 boundary flips}
(see fig.~\ref{figFlipType-1}):
\newline
One adds a vertex to the configuration and connects it to the
two border sites of the {\em type-1} triangle. The connecting
links are called ``virtual'' edges. One then performs a
{\em bulk flip} on the former border link of the {\em type-1} triangle.
Now one selects one of the two virtual boundary links
(each of them with probability one half) and removes it.
The remaining virtual link is then made ``real''.
The so defined operation obviously increases $L$ by 2.
Note that the {\em type-1} flip is allowed only when the resulting
configuration is connected.

\item {\em Type-2 boundary flips}
(see fig.~\ref{figFlipType-2}):
\newline
To the given
{\em type-2} triangle one adds two virtual links as
depicted in fig.~\ref{figFlipType-2}.
Then the {\em type-2} triangle is removed
and, with uniform probability,
one of the two virtual links is selected and removed.
Finally the remaining virtual link is made real.
The {\em type-2 boundary flip} decreases $L$ by 2.
\end{itemize}
In order to be able to prove later the detailed balance
condition (9), the 3 types of flips have to be applied
according to the following prescription.

During the Monte Carlo sweeps one selects again and again
with uniform probability one of the $N$ triangles
as a candidate for updating.

Once a triangle is selected, one
decides what kind of flip will be attempted on it.

Strange as it may look, the constraints of the detailed
balance proof (see below) require that the
kind of the update (flip) is to be
decided {\it a priori}, independent
of the actual type of the triangle on which
it is going to be attempted.

More precisely, one decides
with probability $0 < p < 1$ to perform a {\em boundary flip},
and with probability $1-p$ to perform a {\em bulk flip} ($p$ is
a free parameter which
we chose in our runs to be 0.5).\footnote{Another flip operation,
also consistent with stationarity, is
performing  bulk and
boundary flips in fixed order, e.g., alternating}

If the selected update kind is ``boundary'' and the
triangle is {\em type-0} (``bulk'') then the present updating
attempt is aborted and one proceeds to the selection
of a new candidate triangle.

If the update kind
is ``boundary'' and the triangle is {\em type-1} or
{\em type-2} then one applies the corresponding {\em boundary
flip}.

If the update kind is ``bulk'' one selects
with uniform probability one of the three links (edges) of the
triangle. If the link belongs to the boundary,
the updating attempt is aborted and one proceeds to
the selection of a new triangle. Otherwise the selected
link is bulk flipped.

In Monte-Carlo simulations one must ensure that
the stochastic process is ergodic, i.e., that all
configurations can be reached
starting from an arbitrary initial configuration.

The ergodicity for fixed $N$ and $L$ comes
from the ergodicity of the Alexander (2,2) move (the bulk flip).
Therefore all configurations with fixed $(N,L)$ can be reached by a
finite number of steps. With boundary flips one can get, for
a fixed $N$, from any $L$ to any $\tilde{L}$
provided that $L$ and $\tilde{L}$ have the same parity and that
$3 \leq \tilde{L} \leq N+2$. These two restrictions are specific for the
model that we consider. Thus the combined flips provide an
ergodic process for a fixed number of triangles and a varying boundary
length.

In order to ensure that each configuration
is generated with equal probability we demand that our
Monte Carlo procedure satisfies the detailed balance
condition.
In the absence of matter fields this means that the
probability $P$ to move from one configuration to another
obeys
\be\label {det}
P(\t \ra \tilde{\t}) = P(\tilde{\t} \ra \t) \, .
\ee
It is known that eq.~(\ref{det}) is fulfilled for
bulk flips \cite{closed-DTRS}.
We therefore consider configurations $\tilde{\t}$ that differ from $\t$
by a boundary flip.
The probability that a certain triangle is selected to be boundary flipped
is $\frac{1}{N}$. (Of course, as mentioned above, the triangle will be
actually flipped, only if is {\em not} a {\em type-0} triangle.)
There are four flip possibilities.
In each one a couple of triangles (*,*) from $\t$
(where (*,*) expresses the types of the triangles)
change their type:
\begin{description}
\item [(1+)]
a {\em type-1 boundary flip} changes
a ({\em type-1,type-0}) combination into a
({\em type-2,type-1}) combination.
This operation is applied with probability
$\frac{1}{2N}$. The factor $\frac{1}{2}$ comes from choosing the
neighbour {\em type-0} triangle out of two possible neighbours of the
{\em type-1} triangle.
\item [(1--)]
a {\em type-2 boundary flip} (also selected with probability
$\frac{1}{2N}$) moves
a ({\em type-1,type-2}) combination to a
({\em type-0,type-1}) combination.
This is the move depicted in fig.~\ref{figFlipType-2}.
It is the ``inverse'' operation to (1+).
\item [(2+)]
a {\em type-1 boundary flip} (selected with probability $\frac{1}{N}$)
changes a ({\em type-1,type-1}) into a
({\em type-2 , type-2}) combination. This is the
operation depicted in fig.~\ref{figFlipType-1}.
\item [(2--)]
The ``inverse" operation of (2+) is a
{\em type-2 boundary flip} taking a
({\em type-1,type-2}) combination into a
({\em type-1,type-1}) combination. It is also selected with
probability $\frac{1}{N}$.
\end{description}
The verification of the detailed balance condition is in the statement
that the (+) and (--) operations are inverse to each other and
are selected with the same probability.

Note that our algorithm can be easily adapted to the inclusion
of matter fields (which live on the triangles).

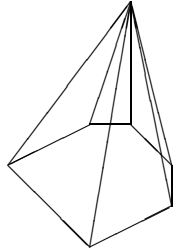
\begin{figure}
\begin{center}
\setlength{\unitlength}{3ex}
\begin{picture}(15,8)(0,0)
 \put(6,0)
 {
  \begin{picture}(8,5)(0,0)
   \put( 1, 0){\line( 2, 1){2}}
   \put( 3, 1){\line( 0, 1){1}}
   \put( 3, 2){\line(-1, 1){1}}
   \put( 2, 3){\line(-1, 0){1}}
   \put( 1, 3){\line(-2,-1){2}}
   \put(-1, 2){\line( 1,-1){2}}
   \put( 1, 0){\line( 1, 6){1}}
   \put( 3, 1){\line(-1, 5){1}}
   \put( 3, 2){\line(-1, 4){1}}
   \put( 2, 3){\line( 0, 1){3}}
   \put( 1, 3){\line( 1, 3){1}}
   \put(-1, 2){\line( 3, 4){3}}
  \end{picture}
 }
\end{picture}
  \parbox[t]{.85\textwidth}
  {
  \caption[The polyhedron start configuration]
  {\label{figPoly}
  Polyhedron start configuration}
  }
\end{center}
\end{figure}

For the actual simulation we used two geometrically very different initial
configurations. This enabled us to
check the ergodicity of our algorithm and also
gave us the possibility to check for critical
slowing down. The two initial configurations were:
\begin{enumerate}
  \item A polyhedron consisting of $N$ triangles, see
   fig.~\ref{figPoly}.
   All triangles are {\em type-1} and share a single vertex.
  \item A configuration that contains 20 {\em type-1}
   triangles and $N-20$ {\em type-0} triangles.
   For large $N$ this initial configuration
   resembles a sphere with a hole in it.
\end{enumerate}

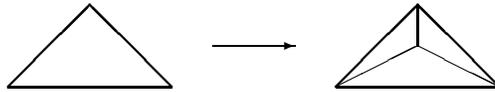
\begin{figure}
\begin{center}
\setlength{\unitlength}{3ex}
\begin{picture}(15,5)(0,0)
 \put(1,1)
 {
  \begin{picture}(8,5)(0,0)
   \thicklines
   \put( 0, 0){\line( 1, 0){4}}
   \put( 0, 0){\line( 1, 1){2}}
   \put( 4, 0){\line(-1, 1){2}}
   \thinlines
   \put( 5, 1){\vector(1,0){2}}
  \end{picture}
 }
 \put(9,1)
 {
  \begin{picture}(8,5)(0,0)
   \thicklines
   \put( 0, 0){\line( 1, 0){4}}
   \put( 0, 0){\line( 1, 1){2}}
   \put( 4, 0){\line(-1, 1){2}}
   \thinlines
   \put( 0, 0){\line( 2, 1){2}}
   \put( 2, 1){\line( 0, 1){1}}
   \put( 2, 1){\line( 2,-1){2}}
  \end{picture}
 }
\end{picture}
  \parbox[t]{.85\textwidth}
  {
  \caption[Growing the initial configuration]
  {\label{figGrow}
  Growing the initial configuration}
  }
\end{center}
\end{figure}

This second initial configuration is constructed by
taking a configuration of the first type with $N=20$
and growing it by inserting vertices in the
middle of the triangles \cite{closed-DTRS},
see fig.~\ref{figGrow}. Each vertex insertion increases $N$ by two.
After each insertion of a vertex, we
performed a sweep.

\section{Results}
\label{SECresults}

Using the algorithm described above we looked at various quantities.
For several different values of $N$ we obtained estimates for
the mean of the boundary length ($\langle L \rangle$), and
the variance of the boundary length
($\langle L^2 \rangle - \langle L \rangle^2$). We also
studied the expectation values of the
number of {\em type-$0,1,2$} triangles and of the number
of triangles $N_f$ that do not have any vertex
on the boundary.

For thermalization we typically performed 15,000 sweeps.
For measurement we typically performed 150,000 sweeps per run
(measuring after each sweep).

$\langle L \rangle$ turns out to be a linear function of
$N$ with high precision. In table~\ref{table:Measured}
we present our Monte Carlo results for $N$ ranging from
$50$ to $6400$. We give always two estimates for $\langle L \rangle$,
one obtained from a run with initial boundary length $L_0 = 20$
and one from a run with $L_0=N$. The results are always nicely
consistent within $1\sigma$ error bars. Fitting the
data with the law $\langle L \rangle = A + B \, N $ we
find $A=1.61(3)$, $B=0.764697(34)$, $\chi^2/{\rm dof}=1.5$ for
the data obtained from runs with $L_0=20$, and
$A=1.62(3)$, $B=0.764643(34)$, $\chi^2/{\rm dof}=2.4$ for the
data obtained from runs with $L_0=N$.

The constant $B$ can be obtained analytically:
The result for our lattice model is
B = $\frac{13}{17}$ = 0.764706 \cite{eti}.
Our fit result from
the $L_0=20$ runs are nicely consistent with this
exact result while the fit to the $L_0=N$ result is consistent
within $2\sigma$.

The linear dependence of $\langle L \rangle $ on $N$ indicates
a fractal dimension of the typical equilibrium configuration
and the fact that it is composed by loosely connected subgraphs.
Further evidence that
this is indeed the situation can be seen in table~\ref{table:Number}
that quotes the expectation values of the number of triangles
not touching the boundary $ \langle N_f \rangle$.
This data indicates
that a typical configuration is tree-shaped
with most links belonging to the boundary.
Most of the branches look like beads of
minimal width (1 link across).

The variance of $L$ goes like $N^{\frac{1}{2}}$. This
suggests that the locations of the further branching within a branch
are random independent variables. In particular, this implies
that there are no long range correlations between the various local
geometric features of these ``surfaces".

\begin{figure}
\begin{center}
\setlength{\unitlength}{0.012500in}%
\begin{picture}(295,300)(70,415)
\thicklines
\put(100,625){\line(-1,-4){ 10}}
\put( 90,585){\line(-4, 3){ 20}}
\put( 70,610){\line( 0,-1){ 20}}
\put( 90,585){\line( 1,-3){ 10}}
\put(100,555){\line(-5, 1){ 25}}
\put( 75,570){\line( 0,-1){ 25}}
\put(100,555){\line( 4,-3){ 40}}
\put(140,525){\line(-1,-4){ 10}}
\put(130,485){\line(-6,-1){ 30}}
\put(100,480){\line(-1, 1){ 15}}
\put(100,480){\line(-1,-2){ 10}}
\put( 90,505){\line(-1,-2){ 10}}
\put( 85,465){\line( 2,-3){ 10}}
\put(130,485){\line( 1,-1){ 25}}
\put(140,525){\line( 1, 0){ 50}}
\put(190,525){\line( 5, 6){ 25}}
\put(215,555){\line( 5,-3){ 25}}
\put(245,550){\line(-1,-2){ 10}}
\put(215,555){\line( 1, 4){ 10}}
\put(225,595){\line( 1, 0){ 45}}
\put(270,595){\line( 3,-5){ 15}}
\put(290,575){\line(-3,-2){ 15}}
\put(270,595){\line( 1, 1){ 30}}
\put(300,625){\line(-3, 5){ 15}}
\put(290,655){\line(-3,-2){ 15}}
\put(300,625){\line( 2,-1){ 40}}
\put(340,605){\line( 1, 2){ 15}}
\put(350,640){\line( 3,-2){ 15}}
\put(340,605){\line( 1,-1){ 20}}
\put(365,590){\line(-3,-2){ 15}}
\put(225,595){\line(-2, 3){ 20}}
\put(205,625){\line(-1, 0){ 25}}
\put(175,635){\line( 1,-4){  5}}
\put(205,625){\line( 1, 4){ 10}}
\put(215,665){\line( 3, 2){ 30}}
\put(200,695){\line( 4,-1){ 40}}
\put(135,425){\line( 2,-1){ 20}}
\put(155,460){\line(-1,-4){ 10}}
\put(190,525){\line( 3,-4){ 15}}
\put(215,505){\line(-4,-1){ 20}}
\put(155,460){\line( 6,-1){ 30}}
\put(185,455){\line( 1, 2){ 10}}
\put(185,455){\line( 4,-5){ 20}}
\put(210,435){\line(-2,-3){ 10}}
\put(190,480){\line( 3,-2){ 15}}
\put(100,625){\line( 6, 1){ 30}}
\put(130,640){\line( 0,-1){ 20}}
\put(100,625){\line(-5, 3){ 25}}
\put( 80,650){\line(-1,-2){ 10}}
\put(215,665){\line(-1, 2){ 25}}
\put(195,715){\line(-1, 0){ 15}}
\put(240,685){\line( 1, 0){ 25}}
\put(265,690){\line( 1,-3){  5}}
\end{picture}
  \parbox[t]{.85\textwidth}
  {
  \caption[Snapshot of a configuration on the dual lattice]
  {\label{figPortrait}
  Snapshot of a configuration on the dual lattice}
  }
\end{center}
\end{figure}
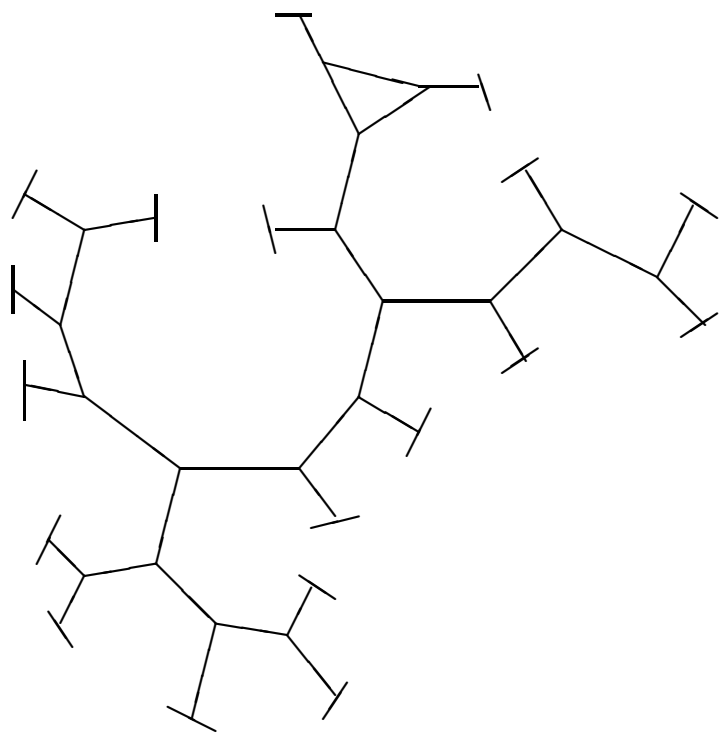

To obtain further information about the global structure of equilibrium
configurations we made some ``snapshots" of small $N$ configurations. The
``portrait''  of such a  typical surface
(represented on the dual lattice) is shown in fig.~\ref{figPortrait}.
This ``snapshot'' substantiates the picture that
the typical surfaces are tree or root like with branching
occurring at random walk steps on the configuration.

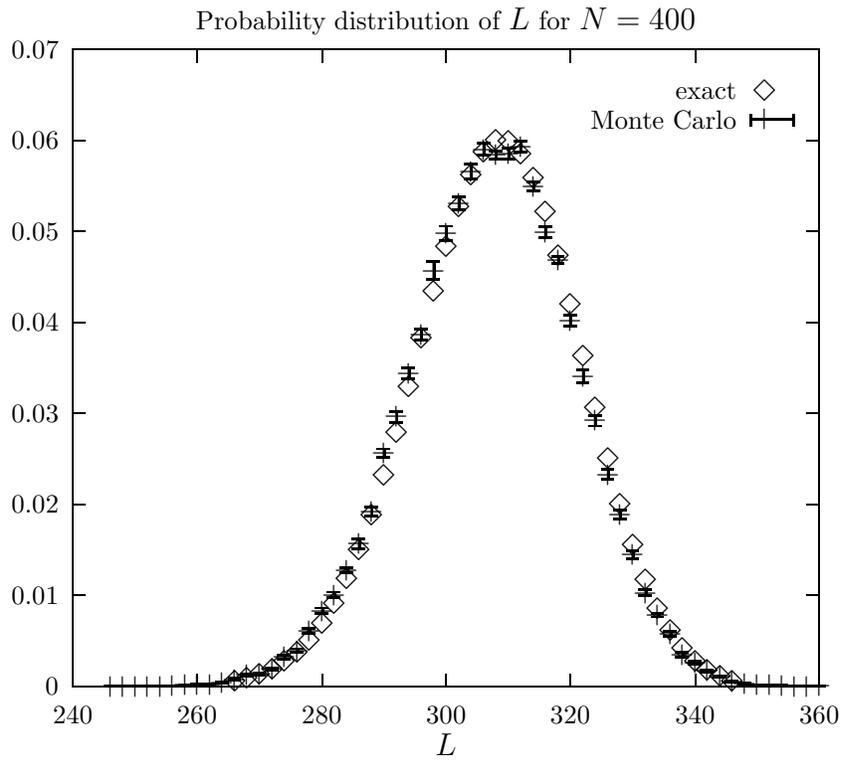
\begin{figure}
\begin{center}
\setlength{\unitlength}{0.240900pt}
\ifx\plotpoint\undefined\newsavebox{\plotpoint}\fi
\sbox{\plotpoint}{\rule[-0.175pt]{0.350pt}{0.350pt}}%
\begin{picture}(1900,1272)(0,0)
\tenrm
\sbox{\plotpoint}{\rule[-0.175pt]{0.350pt}{0.350pt}}%
\put(264,158){\rule[-0.175pt]{282.335pt}{0.350pt}}
\put(264,158){\rule[-0.175pt]{4.818pt}{0.350pt}}
\put(242,158){\makebox(0,0)[r]{0}}
\put(1416,158){\rule[-0.175pt]{4.818pt}{0.350pt}}
\put(264,301){\rule[-0.175pt]{4.818pt}{0.350pt}}
\put(242,301){\makebox(0,0)[r]{0.01}}
\put(1416,301){\rule[-0.175pt]{4.818pt}{0.350pt}}
\put(264,444){\rule[-0.175pt]{4.818pt}{0.350pt}}
\put(242,444){\makebox(0,0)[r]{0.02}}
\put(1416,444){\rule[-0.175pt]{4.818pt}{0.350pt}}
\put(264,587){\rule[-0.175pt]{4.818pt}{0.350pt}}
\put(242,587){\makebox(0,0)[r]{0.03}}
\put(1416,587){\rule[-0.175pt]{4.818pt}{0.350pt}}
\put(264,730){\rule[-0.175pt]{4.818pt}{0.350pt}}
\put(242,730){\makebox(0,0)[r]{0.04}}
\put(1416,730){\rule[-0.175pt]{4.818pt}{0.350pt}}
\put(264,873){\rule[-0.175pt]{4.818pt}{0.350pt}}
\put(242,873){\makebox(0,0)[r]{0.05}}
\put(1416,873){\rule[-0.175pt]{4.818pt}{0.350pt}}
\put(264,1016){\rule[-0.175pt]{4.818pt}{0.350pt}}
\put(242,1016){\makebox(0,0)[r]{0.06}}
\put(1416,1016){\rule[-0.175pt]{4.818pt}{0.350pt}}
\put(264,1159){\rule[-0.175pt]{4.818pt}{0.350pt}}
\put(242,1159){\makebox(0,0)[r]{0.07}}
\put(1416,1159){\rule[-0.175pt]{4.818pt}{0.350pt}}
\put(264,158){\rule[-0.175pt]{0.350pt}{4.818pt}}
\put(264,113){\makebox(0,0){240}}
\put(264,1139){\rule[-0.175pt]{0.350pt}{4.818pt}}
\put(459,158){\rule[-0.175pt]{0.350pt}{4.818pt}}
\put(459,113){\makebox(0,0){260}}
\put(459,1139){\rule[-0.175pt]{0.350pt}{4.818pt}}
\put(655,158){\rule[-0.175pt]{0.350pt}{4.818pt}}
\put(655,113){\makebox(0,0){280}}
\put(655,1139){\rule[-0.175pt]{0.350pt}{4.818pt}}
\put(850,158){\rule[-0.175pt]{0.350pt}{4.818pt}}
\put(850,113){\makebox(0,0){300}}
\put(850,1139){\rule[-0.175pt]{0.350pt}{4.818pt}}
\put(1045,158){\rule[-0.175pt]{0.350pt}{4.818pt}}
\put(1045,113){\makebox(0,0){320}}
\put(1045,1139){\rule[-0.175pt]{0.350pt}{4.818pt}}
\put(1241,158){\rule[-0.175pt]{0.350pt}{4.818pt}}
\put(1241,113){\makebox(0,0){340}}
\put(1241,1139){\rule[-0.175pt]{0.350pt}{4.818pt}}
\put(1436,158){\rule[-0.175pt]{0.350pt}{4.818pt}}
\put(1436,113){\makebox(0,0){360}}
\put(1436,1139){\rule[-0.175pt]{0.350pt}{4.818pt}}
\put(264,158){\rule[-0.175pt]{282.335pt}{0.350pt}}
\put(1436,158){\rule[-0.175pt]{0.350pt}{241.141pt}}
\put(264,1159){\rule[-0.175pt]{282.335pt}{0.350pt}}
\put(850,68){\makebox(0,0){$L$}}
\put(850,1204){\makebox(0,0){Probability distribution of $L$ for $N=400$}}
\put(264,158){\rule[-0.175pt]{0.350pt}{241.141pt}}
\put(1306,1094){\makebox(0,0)[r]{exact}}
\put(1350,1094){\raisebox{-1.2pt}{\makebox(0,0){$\Diamond$}}}
\put(518,167){\raisebox{-1.2pt}{\makebox(0,0){$\Diamond$}}}
\put(537,171){\raisebox{-1.2pt}{\makebox(0,0){$\Diamond$}}}
\put(557,177){\raisebox{-1.2pt}{\makebox(0,0){$\Diamond$}}}
\put(577,185){\raisebox{-1.2pt}{\makebox(0,0){$\Diamond$}}}
\put(596,197){\raisebox{-1.2pt}{\makebox(0,0){$\Diamond$}}}
\put(616,211){\raisebox{-1.2pt}{\makebox(0,0){$\Diamond$}}}
\put(635,231){\raisebox{-1.2pt}{\makebox(0,0){$\Diamond$}}}
\put(655,257){\raisebox{-1.2pt}{\makebox(0,0){$\Diamond$}}}
\put(674,288){\raisebox{-1.2pt}{\makebox(0,0){$\Diamond$}}}
\put(694,327){\raisebox{-1.2pt}{\makebox(0,0){$\Diamond$}}}
\put(713,373){\raisebox{-1.2pt}{\makebox(0,0){$\Diamond$}}}
\put(733,427){\raisebox{-1.2pt}{\makebox(0,0){$\Diamond$}}}
\put(752,489){\raisebox{-1.2pt}{\makebox(0,0){$\Diamond$}}}
\put(772,557){\raisebox{-1.2pt}{\makebox(0,0){$\Diamond$}}}
\put(791,629){\raisebox{-1.2pt}{\makebox(0,0){$\Diamond$}}}
\put(811,705){\raisebox{-1.2pt}{\makebox(0,0){$\Diamond$}}}
\put(830,778){\raisebox{-1.2pt}{\makebox(0,0){$\Diamond$}}}
\put(850,849){\raisebox{-1.2pt}{\makebox(0,0){$\Diamond$}}}
\put(870,911){\raisebox{-1.2pt}{\makebox(0,0){$\Diamond$}}}
\put(889,962){\raisebox{-1.2pt}{\makebox(0,0){$\Diamond$}}}
\put(909,998){\raisebox{-1.2pt}{\makebox(0,0){$\Diamond$}}}
\put(928,1016){\raisebox{-1.2pt}{\makebox(0,0){$\Diamond$}}}
\put(948,1015){\raisebox{-1.2pt}{\makebox(0,0){$\Diamond$}}}
\put(967,995){\raisebox{-1.2pt}{\makebox(0,0){$\Diamond$}}}
\put(987,957){\raisebox{-1.2pt}{\makebox(0,0){$\Diamond$}}}
\put(1006,903){\raisebox{-1.2pt}{\makebox(0,0){$\Diamond$}}}
\put(1026,835){\raisebox{-1.2pt}{\makebox(0,0){$\Diamond$}}}
\put(1045,759){\raisebox{-1.2pt}{\makebox(0,0){$\Diamond$}}}
\put(1065,677){\raisebox{-1.2pt}{\makebox(0,0){$\Diamond$}}}
\put(1084,596){\raisebox{-1.2pt}{\makebox(0,0){$\Diamond$}}}
\put(1104,517){\raisebox{-1.2pt}{\makebox(0,0){$\Diamond$}}}
\put(1123,445){\raisebox{-1.2pt}{\makebox(0,0){$\Diamond$}}}
\put(1143,380){\raisebox{-1.2pt}{\makebox(0,0){$\Diamond$}}}
\put(1163,326){\raisebox{-1.2pt}{\makebox(0,0){$\Diamond$}}}
\put(1182,281){\raisebox{-1.2pt}{\makebox(0,0){$\Diamond$}}}
\put(1202,246){\raisebox{-1.2pt}{\makebox(0,0){$\Diamond$}}}
\put(1221,218){\raisebox{-1.2pt}{\makebox(0,0){$\Diamond$}}}
\put(1241,198){\raisebox{-1.2pt}{\makebox(0,0){$\Diamond$}}}
\put(1260,184){\raisebox{-1.2pt}{\makebox(0,0){$\Diamond$}}}
\put(1280,174){\raisebox{-1.2pt}{\makebox(0,0){$\Diamond$}}}
\put(1299,167){\raisebox{-1.2pt}{\makebox(0,0){$\Diamond$}}}
\sbox{\plotpoint}{\rule[-0.350pt]{0.700pt}{0.700pt}}%
\put(1306,1049){\makebox(0,0)[r]{Monte Carlo}}
\put(1350,1049){\makebox(0,0){$+$}}
\put(323,158){\makebox(0,0){$+$}}
\put(342,158){\makebox(0,0){$+$}}
\put(362,158){\makebox(0,0){$+$}}
\put(381,158){\makebox(0,0){$+$}}
\put(401,158){\makebox(0,0){$+$}}
\put(420,159){\makebox(0,0){$+$}}
\put(440,160){\makebox(0,0){$+$}}
\put(459,160){\makebox(0,0){$+$}}
\put(479,161){\makebox(0,0){$+$}}
\put(498,164){\makebox(0,0){$+$}}
\put(518,169){\makebox(0,0){$+$}}
\put(537,176){\makebox(0,0){$+$}}
\put(557,177){\makebox(0,0){$+$}}
\put(577,185){\makebox(0,0){$+$}}
\put(596,204){\makebox(0,0){$+$}}
\put(616,214){\makebox(0,0){$+$}}
\put(635,245){\makebox(0,0){$+$}}
\put(655,277){\makebox(0,0){$+$}}
\put(674,301){\makebox(0,0){$+$}}
\put(694,340){\makebox(0,0){$+$}}
\put(713,382){\makebox(0,0){$+$}}
\put(733,432){\makebox(0,0){$+$}}
\put(752,524){\makebox(0,0){$+$}}
\put(772,582){\makebox(0,0){$+$}}
\put(791,650){\makebox(0,0){$+$}}
\put(811,711){\makebox(0,0){$+$}}
\put(830,811){\makebox(0,0){$+$}}
\put(850,870){\makebox(0,0){$+$}}
\put(870,917){\makebox(0,0){$+$}}
\put(889,967){\makebox(0,0){$+$}}
\put(909,1002){\makebox(0,0){$+$}}
\put(928,993){\makebox(0,0){$+$}}
\put(948,995){\makebox(0,0){$+$}}
\put(967,1006){\makebox(0,0){$+$}}
\put(987,944){\makebox(0,0){$+$}}
\put(1006,871){\makebox(0,0){$+$}}
\put(1026,828){\makebox(0,0){$+$}}
\put(1045,733){\makebox(0,0){$+$}}
\put(1065,645){\makebox(0,0){$+$}}
\put(1084,576){\makebox(0,0){$+$}}
\put(1104,491){\makebox(0,0){$+$}}
\put(1123,428){\makebox(0,0){$+$}}
\put(1143,365){\makebox(0,0){$+$}}
\put(1163,305){\makebox(0,0){$+$}}
\put(1182,270){\makebox(0,0){$+$}}
\put(1202,241){\makebox(0,0){$+$}}
\put(1221,208){\makebox(0,0){$+$}}
\put(1241,195){\makebox(0,0){$+$}}
\put(1260,182){\makebox(0,0){$+$}}
\put(1280,173){\makebox(0,0){$+$}}
\put(1299,166){\makebox(0,0){$+$}}
\put(1319,162){\makebox(0,0){$+$}}
\put(1338,160){\makebox(0,0){$+$}}
\put(1358,160){\makebox(0,0){$+$}}
\put(1377,159){\makebox(0,0){$+$}}
\put(1397,158){\makebox(0,0){$+$}}
\put(1416,158){\makebox(0,0){$+$}}
\put(1436,159){\makebox(0,0){$+$}}
\put(1328,1049){\rule[-0.350pt]{15.899pt}{0.700pt}}
\put(1328,1039){\rule[-0.350pt]{0.700pt}{4.818pt}}
\put(1394,1039){\rule[-0.350pt]{0.700pt}{4.818pt}}
\put(323,158){\usebox{\plotpoint}}
\put(313,158){\rule[-0.350pt]{4.818pt}{0.700pt}}
\put(313,158){\rule[-0.350pt]{4.818pt}{0.700pt}}
\put(342,158){\usebox{\plotpoint}}
\put(332,158){\rule[-0.350pt]{4.818pt}{0.700pt}}
\put(332,159){\rule[-0.350pt]{4.818pt}{0.700pt}}
\put(362,158){\usebox{\plotpoint}}
\put(352,158){\rule[-0.350pt]{4.818pt}{0.700pt}}
\put(352,158){\rule[-0.350pt]{4.818pt}{0.700pt}}
\put(381,158){\usebox{\plotpoint}}
\put(371,158){\rule[-0.350pt]{4.818pt}{0.700pt}}
\put(371,159){\rule[-0.350pt]{4.818pt}{0.700pt}}
\put(401,158){\usebox{\plotpoint}}
\put(391,158){\rule[-0.350pt]{4.818pt}{0.700pt}}
\put(391,159){\rule[-0.350pt]{4.818pt}{0.700pt}}
\put(420,158){\usebox{\plotpoint}}
\put(410,158){\rule[-0.350pt]{4.818pt}{0.700pt}}
\put(410,159){\rule[-0.350pt]{4.818pt}{0.700pt}}
\put(440,160){\usebox{\plotpoint}}
\put(430,160){\rule[-0.350pt]{4.818pt}{0.700pt}}
\put(430,160){\rule[-0.350pt]{4.818pt}{0.700pt}}
\put(459,160){\usebox{\plotpoint}}
\put(449,160){\rule[-0.350pt]{4.818pt}{0.700pt}}
\put(449,161){\rule[-0.350pt]{4.818pt}{0.700pt}}
\put(479,161){\usebox{\plotpoint}}
\put(469,161){\rule[-0.350pt]{4.818pt}{0.700pt}}
\put(469,162){\rule[-0.350pt]{4.818pt}{0.700pt}}
\put(498,163){\usebox{\plotpoint}}
\put(488,163){\rule[-0.350pt]{4.818pt}{0.700pt}}
\put(488,165){\rule[-0.350pt]{4.818pt}{0.700pt}}
\put(518,168){\rule[-0.350pt]{0.700pt}{0.723pt}}
\put(508,168){\rule[-0.350pt]{4.818pt}{0.700pt}}
\put(508,171){\rule[-0.350pt]{4.818pt}{0.700pt}}
\put(537,174){\rule[-0.350pt]{0.700pt}{0.723pt}}
\put(527,174){\rule[-0.350pt]{4.818pt}{0.700pt}}
\put(527,177){\rule[-0.350pt]{4.818pt}{0.700pt}}
\put(557,175){\rule[-0.350pt]{0.700pt}{0.964pt}}
\put(547,175){\rule[-0.350pt]{4.818pt}{0.700pt}}
\put(547,179){\rule[-0.350pt]{4.818pt}{0.700pt}}
\put(577,183){\rule[-0.350pt]{0.700pt}{0.964pt}}
\put(567,183){\rule[-0.350pt]{4.818pt}{0.700pt}}
\put(567,187){\rule[-0.350pt]{4.818pt}{0.700pt}}
\put(596,201){\rule[-0.350pt]{0.700pt}{1.445pt}}
\put(586,201){\rule[-0.350pt]{4.818pt}{0.700pt}}
\put(586,207){\rule[-0.350pt]{4.818pt}{0.700pt}}
\put(616,212){\rule[-0.350pt]{0.700pt}{1.204pt}}
\put(606,212){\rule[-0.350pt]{4.818pt}{0.700pt}}
\put(606,217){\rule[-0.350pt]{4.818pt}{0.700pt}}
\put(635,241){\rule[-0.350pt]{0.700pt}{1.927pt}}
\put(625,241){\rule[-0.350pt]{4.818pt}{0.700pt}}
\put(625,249){\rule[-0.350pt]{4.818pt}{0.700pt}}
\put(655,273){\rule[-0.350pt]{0.700pt}{1.927pt}}
\put(645,273){\rule[-0.350pt]{4.818pt}{0.700pt}}
\put(645,281){\rule[-0.350pt]{4.818pt}{0.700pt}}
\put(674,297){\rule[-0.350pt]{0.700pt}{2.168pt}}
\put(664,297){\rule[-0.350pt]{4.818pt}{0.700pt}}
\put(664,306){\rule[-0.350pt]{4.818pt}{0.700pt}}
\put(694,336){\rule[-0.350pt]{0.700pt}{2.168pt}}
\put(684,336){\rule[-0.350pt]{4.818pt}{0.700pt}}
\put(684,345){\rule[-0.350pt]{4.818pt}{0.700pt}}
\put(713,374){\rule[-0.350pt]{0.700pt}{3.854pt}}
\put(703,374){\rule[-0.350pt]{4.818pt}{0.700pt}}
\put(703,390){\rule[-0.350pt]{4.818pt}{0.700pt}}
\put(733,426){\rule[-0.350pt]{0.700pt}{3.132pt}}
\put(723,426){\rule[-0.350pt]{4.818pt}{0.700pt}}
\put(723,439){\rule[-0.350pt]{4.818pt}{0.700pt}}
\put(752,518){\rule[-0.350pt]{0.700pt}{3.132pt}}
\put(742,518){\rule[-0.350pt]{4.818pt}{0.700pt}}
\put(742,531){\rule[-0.350pt]{4.818pt}{0.700pt}}
\put(772,573){\rule[-0.350pt]{0.700pt}{4.095pt}}
\put(762,573){\rule[-0.350pt]{4.818pt}{0.700pt}}
\put(762,590){\rule[-0.350pt]{4.818pt}{0.700pt}}
\put(791,641){\rule[-0.350pt]{0.700pt}{4.336pt}}
\put(781,641){\rule[-0.350pt]{4.818pt}{0.700pt}}
\put(781,659){\rule[-0.350pt]{4.818pt}{0.700pt}}
\put(811,702){\rule[-0.350pt]{0.700pt}{4.336pt}}
\put(801,702){\rule[-0.350pt]{4.818pt}{0.700pt}}
\put(801,720){\rule[-0.350pt]{4.818pt}{0.700pt}}
\put(830,797){\rule[-0.350pt]{0.700pt}{6.745pt}}
\put(820,797){\rule[-0.350pt]{4.818pt}{0.700pt}}
\put(820,825){\rule[-0.350pt]{4.818pt}{0.700pt}}
\put(850,859){\rule[-0.350pt]{0.700pt}{5.300pt}}
\put(840,859){\rule[-0.350pt]{4.818pt}{0.700pt}}
\put(840,881){\rule[-0.350pt]{4.818pt}{0.700pt}}
\put(870,907){\rule[-0.350pt]{0.700pt}{4.818pt}}
\put(860,907){\rule[-0.350pt]{4.818pt}{0.700pt}}
\put(860,927){\rule[-0.350pt]{4.818pt}{0.700pt}}
\put(889,955){\rule[-0.350pt]{0.700pt}{5.782pt}}
\put(879,955){\rule[-0.350pt]{4.818pt}{0.700pt}}
\put(879,979){\rule[-0.350pt]{4.818pt}{0.700pt}}
\put(909,993){\rule[-0.350pt]{0.700pt}{4.336pt}}
\put(899,993){\rule[-0.350pt]{4.818pt}{0.700pt}}
\put(899,1011){\rule[-0.350pt]{4.818pt}{0.700pt}}
\put(928,986){\rule[-0.350pt]{0.700pt}{3.132pt}}
\put(918,986){\rule[-0.350pt]{4.818pt}{0.700pt}}
\put(918,999){\rule[-0.350pt]{4.818pt}{0.700pt}}
\put(948,986){\rule[-0.350pt]{0.700pt}{4.336pt}}
\put(938,986){\rule[-0.350pt]{4.818pt}{0.700pt}}
\put(938,1004){\rule[-0.350pt]{4.818pt}{0.700pt}}
\put(967,997){\rule[-0.350pt]{0.700pt}{4.336pt}}
\put(957,997){\rule[-0.350pt]{4.818pt}{0.700pt}}
\put(957,1015){\rule[-0.350pt]{4.818pt}{0.700pt}}
\put(987,937){\rule[-0.350pt]{0.700pt}{3.373pt}}
\put(977,937){\rule[-0.350pt]{4.818pt}{0.700pt}}
\put(977,951){\rule[-0.350pt]{4.818pt}{0.700pt}}
\put(1006,863){\rule[-0.350pt]{0.700pt}{4.095pt}}
\put(996,863){\rule[-0.350pt]{4.818pt}{0.700pt}}
\put(996,880){\rule[-0.350pt]{4.818pt}{0.700pt}}
\put(1026,822){\rule[-0.350pt]{0.700pt}{2.891pt}}
\put(1016,822){\rule[-0.350pt]{4.818pt}{0.700pt}}
\put(1016,834){\rule[-0.350pt]{4.818pt}{0.700pt}}
\put(1045,724){\rule[-0.350pt]{0.700pt}{4.095pt}}
\put(1035,724){\rule[-0.350pt]{4.818pt}{0.700pt}}
\put(1035,741){\rule[-0.350pt]{4.818pt}{0.700pt}}
\put(1065,635){\rule[-0.350pt]{0.700pt}{4.818pt}}
\put(1055,635){\rule[-0.350pt]{4.818pt}{0.700pt}}
\put(1055,655){\rule[-0.350pt]{4.818pt}{0.700pt}}
\put(1084,567){\rule[-0.350pt]{0.700pt}{4.095pt}}
\put(1074,567){\rule[-0.350pt]{4.818pt}{0.700pt}}
\put(1074,584){\rule[-0.350pt]{4.818pt}{0.700pt}}
\put(1104,484){\rule[-0.350pt]{0.700pt}{3.613pt}}
\put(1094,484){\rule[-0.350pt]{4.818pt}{0.700pt}}
\put(1094,499){\rule[-0.350pt]{4.818pt}{0.700pt}}
\put(1123,421){\rule[-0.350pt]{0.700pt}{3.373pt}}
\put(1113,421){\rule[-0.350pt]{4.818pt}{0.700pt}}
\put(1113,435){\rule[-0.350pt]{4.818pt}{0.700pt}}
\put(1143,359){\rule[-0.350pt]{0.700pt}{2.891pt}}
\put(1133,359){\rule[-0.350pt]{4.818pt}{0.700pt}}
\put(1133,371){\rule[-0.350pt]{4.818pt}{0.700pt}}
\put(1163,300){\rule[-0.350pt]{0.700pt}{2.409pt}}
\put(1153,300){\rule[-0.350pt]{4.818pt}{0.700pt}}
\put(1153,310){\rule[-0.350pt]{4.818pt}{0.700pt}}
\put(1182,267){\rule[-0.350pt]{0.700pt}{1.204pt}}
\put(1172,267){\rule[-0.350pt]{4.818pt}{0.700pt}}
\put(1172,272){\rule[-0.350pt]{4.818pt}{0.700pt}}
\put(1202,237){\rule[-0.350pt]{0.700pt}{1.927pt}}
\put(1192,237){\rule[-0.350pt]{4.818pt}{0.700pt}}
\put(1192,245){\rule[-0.350pt]{4.818pt}{0.700pt}}
\put(1221,204){\rule[-0.350pt]{0.700pt}{1.686pt}}
\put(1211,204){\rule[-0.350pt]{4.818pt}{0.700pt}}
\put(1211,211){\rule[-0.350pt]{4.818pt}{0.700pt}}
\put(1241,193){\rule[-0.350pt]{0.700pt}{0.964pt}}
\put(1231,193){\rule[-0.350pt]{4.818pt}{0.700pt}}
\put(1231,197){\rule[-0.350pt]{4.818pt}{0.700pt}}
\put(1260,180){\rule[-0.350pt]{0.700pt}{0.964pt}}
\put(1250,180){\rule[-0.350pt]{4.818pt}{0.700pt}}
\put(1250,184){\rule[-0.350pt]{4.818pt}{0.700pt}}
\put(1280,172){\usebox{\plotpoint}}
\put(1270,172){\rule[-0.350pt]{4.818pt}{0.700pt}}
\put(1270,174){\rule[-0.350pt]{4.818pt}{0.700pt}}
\put(1299,165){\usebox{\plotpoint}}
\put(1289,165){\rule[-0.350pt]{4.818pt}{0.700pt}}
\put(1289,167){\rule[-0.350pt]{4.818pt}{0.700pt}}
\put(1319,162){\usebox{\plotpoint}}
\put(1309,162){\rule[-0.350pt]{4.818pt}{0.700pt}}
\put(1309,163){\rule[-0.350pt]{4.818pt}{0.700pt}}
\put(1338,159){\usebox{\plotpoint}}
\put(1328,159){\rule[-0.350pt]{4.818pt}{0.700pt}}
\put(1328,160){\rule[-0.350pt]{4.818pt}{0.700pt}}
\put(1358,159){\usebox{\plotpoint}}
\put(1348,159){\rule[-0.350pt]{4.818pt}{0.700pt}}
\put(1348,160){\rule[-0.350pt]{4.818pt}{0.700pt}}
\put(1377,159){\usebox{\plotpoint}}
\put(1367,159){\rule[-0.350pt]{4.818pt}{0.700pt}}
\put(1367,160){\rule[-0.350pt]{4.818pt}{0.700pt}}
\put(1397,158){\usebox{\plotpoint}}
\put(1387,158){\rule[-0.350pt]{4.818pt}{0.700pt}}
\put(1387,159){\rule[-0.350pt]{4.818pt}{0.700pt}}
\put(1416,158){\usebox{\plotpoint}}
\put(1406,158){\rule[-0.350pt]{4.818pt}{0.700pt}}
\put(1406,158){\rule[-0.350pt]{4.818pt}{0.700pt}}
\put(1436,158){\usebox{\plotpoint}}
\put(1426,158){\rule[-0.350pt]{4.818pt}{0.700pt}}
\put(1426,159){\rule[-0.350pt]{4.818pt}{0.700pt}}
\end{picture}
  \parbox[t]{.85\textwidth}
  {
  \caption[Probability distribution of $L$ for $N=400$]
  {\label{figWaveFunction}
  Monte Carlo (points with error bars) and exact results (diamonds)
  for the wave function $\psi(L)$ for $N=400$}
  }
\end{center}
\end{figure}

Fig.~\ref{figWaveFunction} shows an example for the statistical distribution
of the border length $L$ for $N=400$ (the wave function $\psi(L)$).
The diamonds show the exact result \cite{eti}, the points with
error bars show estimates produced with our algorithm. The
error bars were obtained as follows: The total of 100,000
measurements of $L$ was produced in 10 independent runs, each of
size 10,000 sweeps. The sweeps consisted of 10,000 flips
of each type.
A histogram was determined for each subsample separately.
The error bars were obtained as the $1\sigma$ error from the
10 independent measurements of the histogram.

In conclusion we have presented results from a simulation study
 of pure 2-D quantum
gravity on open DTRS's. The numerical results
agree with results obtained by analytical calculations \cite{eti}.
In addition our simulation algorithm also allowed us to study the geometry
of the typical surface configurations.

Our results encourage further advance
by adding matter fields.

\section*{Acknowledgments}
We would like to thank S. Elitzur, T. Mohaupt, T.~Wittlich and
E.~Rabinovici for stimulating discussions.

\newpage

\listoftables

\listoffigures

\newpage

\begin{figure}
\begin{center}
\setlength{\unitlength}{3ex}

\begin{picture}(15,7)(0,0)
 \put(0,3)
 {
  \begin{picture}(8,5)(0,0)
   \put( 0, -3){(a)}
   \put( 2, 0){\line( 1, 1){1}}
   \put( 3, 1){\line(-2,-3){2}}
   \thicklines
   \put( 0, 0){\line( 1, 0){2}}
   \put( 0, 0){\line( 1,-2){1}}
   \put( 2, 0){\line(-1,-2){1}}
   \thinlines
   \put(-3, 1){\line( 4,-3){4}}
   \put(-3, 1){\line( 3,-1){3}}
   \multiput(0, 0)(0.1,0.2){10}{\makebox(0,0){.}}
   \multiput(2, 0)(-0.1,0.2){10}{\makebox(0,0){.}}
   \put(1,2){\makebox(0,0){$\bullet$}}
  \end{picture}
 }
 \put(7,3)
 {
  \begin{picture}(8,5)(0,0)
   \put( 0, -3){(b)}
   \put( 2, 0){\line( 1, 1){1}}
   \put( 3, 1){\line(-2,-3){2}}
   \put( 0, 0){\line( 1,-2){1}}
   \put( 2, 0){\line(-1,-2){1}}
   \put(-3, 1){\line( 4,-3){4}}
   \put(-3, 1){\line( 3,-1){3}}
   \multiput(1,-2)(0.,0.2){20}{\makebox(0,0){.}}
   \multiput(0, 0)(0.1,0.2){10}{\makebox(0,0){.}}
   \multiput(2, 0)(-0.1,0.2){10}{\makebox(0,0){.}}
   \put(1,2){\makebox(0,0){$\bullet$}}
  \end{picture}
 }
 \put(14,3)
 {
  \begin{picture}(8,5)(0,0)
   \put( 0, -3){(c)}
   \put( 2, 0){\line( 1, 1){1}}
   \put( 3, 1){\line(-2,-3){2}}
   \put( 0, 0){\line( 1,-2){1}}
   \put( 2, 0){\line(-1,-2){1}}
   \put(-3, 1){\line( 4,-3){4}}
   \put(-3, 1){\line( 3,-1){3}}
   \put( 0, 0){\line( 1, 2){1}}
   \put( 1,-2){\line( 0, 1){4}}
  \end{picture}
 }
\end{picture}
  \parbox[t]{.85\textwidth}
  {
  \caption[The {\em type-1} boundary flip]
  {\label{figFlipType-1}
  How the {\em type-1 boundary flip} works: (a) A vertex is added to the
  configuration and connected to the two border sites of the {\em type-1}
  triangle. (b) A {bulk flip} is performed. (c) One of the two virtual
  boundary edges is removed. The remaining edges are made real.}
  }
\end{center}
\end{figure}
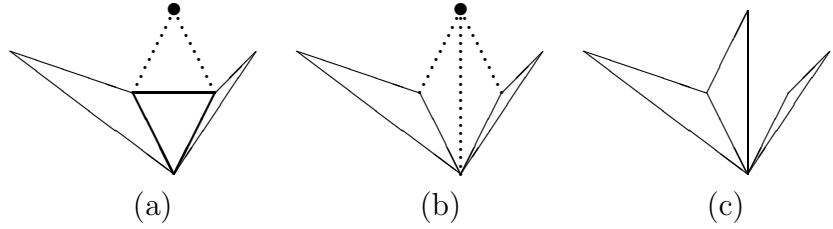

\begin{table}
\begin{center}
\begin{tabular}{|r|r|r|}
\hline
\mc{1}{|c}{$N$}   &
\mc{1}{|c}{$L_0$} & \mc{1}{|c|}{$\langle L \rangle$}     \\
\hline
   50   &   20   &    39.83(5)    \\
   50   &   50   &    39.78(5)    \\
  100   &   20   &    78.05(5)    \\
  100   &  100   &    78.11(5)    \\
  200   &   20   &    154.54(6)   \\
  200   &  200   &    154.63(6)   \\
  400   &   20   &    307.62(8)   \\
  400   &  400   &    307.45(8)   \\
  800   &   20   &    613.4(1)     \\
  800   &  800   &    613.4(1)      \\
  1600  &   20   &   1225.1(1)      \\
  1600  &  1600  &   1224.9(1)      \\
  3200  &   20   &   2448.7(2)      \\
  3200  &  3200  &   2448.7(2)      \\
  6400  &    20  &   4895.6(3)      \\
  6400  &  6400  &   4895.4(3)      \\
\hline
 \end{tabular}
  \parbox[t]{.85\textwidth}
  {
  \caption[Monte Carlo estimates for $\langle L \rangle$]
  {\label{table:Measured}
  Monte Carlo estimates for $\langle L \rangle$}
  }
\end{center}
\end{table}

\begin{table}
\begin{center}
\begin{tabular}{|r|r|}
\hline
\mc{1}{|c}{$N$} & \mc{1}{|c|}{$\langle N_f \rangle $} \\
\hline
  50    &   0.51(1)  \\
 100    &   0.96(1)  \\
 200    &   1.92(2)  \\
 400    &   3.85(2)  \\
 800    &   7.67(3)  \\
1600    &  15.27(5)  \\
3200    &  30.47(6)  \\
\hline
 \end{tabular}
  \parbox[t]{.85\textwidth}
  {
  \caption[Monte Carlo estimates for $\langle N_f \rangle$]
  {\label{table:Number}
  Monte Carlo estimates for $\langle N_f \rangle$}
  }
\end{center}
\end{table}

\end{document}